\documentclass[prl,footinbib,twocolumn,preprintnumbers,amsmath,longbibliography,amssymb]{revtex4-1}

\usepackage{amssymb}
\usepackage{amsmath}

\usepackage{color}
\usepackage[pdftex]{graphicx}

\usepackage{empheq}
\usepackage{mathtools}

\usepackage{makecell}
\usepackage{subfigure}
\usepackage{enumerate}
\usepackage{mathtools}
\usepackage{stmaryrd}
\usepackage{enumitem}
\usepackage{textcomp}
\usepackage{gensymb}

\usepackage{natbib}

\newcommand{\ex}[1]{\mathrm{e}^{#1}}

\newcommand{\dd}[0]{\mathrm{d}}

\newcommand{\rr}[0]{\boldsymbol{r}}

\newcommand{\kB}[0]{k_{\mathrm{B}}}

\newcommand{\cC}{\mathcal{C}}

\definecolor{darkblue}{rgb}{0,0,0.6}
\definecolor{darkred}{rgb}{0.6,0,0}
\usepackage[colorlinks=true,urlcolor=darkblue,citecolor=darkblue,linkcolor=darkred]{hyperref}

\begin{document}

\title{Microscopic and stochastic simulations of chemically active droplets}

\author{Roxanne Berthin}
\thanks{These two authors contributed equally.}
\affiliation{Sorbonne Universit\'e, CNRS, Physico-Chimie des \'Electrolytes et Nanosyst\`emes Interfaciaux (PHENIX), 4 Place Jussieu, 75005 Paris, France}

\author{Jacques Fries}
\thanks{These two authors contributed equally.}
\affiliation{Sorbonne Universit\'e, CNRS, Physico-Chimie des \'Electrolytes et Nanosyst\`emes Interfaciaux (PHENIX), 4 Place Jussieu, 75005 Paris, France}

\author{Marie Jardat}
\affiliation{Sorbonne Universit\'e, CNRS, Physico-Chimie des \'Electrolytes et Nanosyst\`emes Interfaciaux (PHENIX), 4 Place Jussieu, 75005 Paris, France}

\author{Vincent Dahirel}
\affiliation{Sorbonne Universit\'e, CNRS, Physico-Chimie des \'Electrolytes et Nanosyst\`emes Interfaciaux (PHENIX), 4 Place Jussieu, 75005 Paris, France}

\author{Pierre Illien}
\affiliation{Sorbonne Universit\'e, CNRS, Physico-Chimie des \'Electrolytes et Nanosyst\`emes Interfaciaux (PHENIX), 4 Place Jussieu, 75005 Paris, France}

\begin{abstract}
Biomolecular condensates play a central role in the spatial organization of living matter. Their formation is now well understood as a form of liquid-liquid phase separation that occurs very far from equilibrium. For instance, they can be modeled as active droplets, where the combination of molecular interactions and chemical reactions result in microphase separation. However, so far, models of chemically active droplets are spatially continuous and deterministic. Therefore, the relationship between the microscopic parameters of the models and some crucial properties of active droplets (such as their polydispersity, their shape anisotropy, or their typical lifetime) is yet to be established. In this work, we address this question computationally, using Brownian dynamics simulations of chemically active droplets:  the building blocks are represented explicitly as particles that interact with attractive or repulsive interactions, depending on whether they are in a droplet-forming state or not. Thanks to this microscopic and stochastic view of the problem, we reveal how driving the system away from equilibrium in a controlled way determines the fluctuations and dynamics of active emulsions.
\end{abstract}

\date{\today}

\maketitle

\emph{Introduction.---} The formation of biomolecular condensates is a central feature of the spatial organization of living matter at the subcellular and subnuclear levels, and plays a key role in the regulation of multiple metabolic processes~\cite{Banani2017, Shin2017}. During recent years, a significant research effort has been devoted to understanding the physical and chemical mechanisms that govern the formation and the dynamics of these membrane-less organelles, both from experimental and theoretical point of views. The now widely spread picture is that of a form of liquid-liquid phase separation that takes place very far from thermodynamic equilibrium, and that typically results in the selection of a well-defined size for the condensates~\cite{Hyman2014,Abbas2021,Gouveia2022,Soding2020}.

\begin{figure}[b]
    \centering
    \includegraphics[width=\columnwidth]{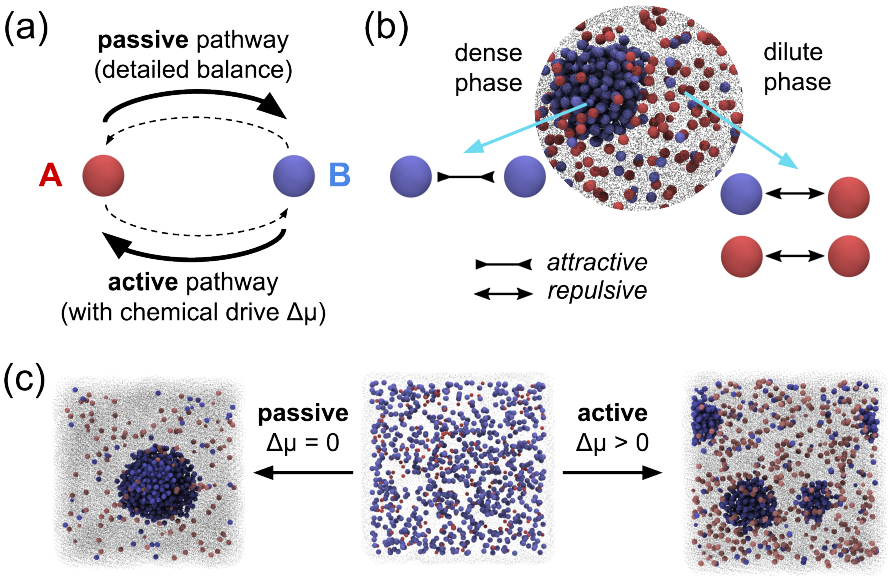}
\caption{(a) Two chemical pathways for the interconversions  $A \rightleftharpoons B$: a passive pathway, which fulfils detailed balance, and an active pathway, which violates it, because of a constant chemical drive $\Delta \mu$. (b) System under study:  chemically active species ($A$ and $B$) and inert particles ($C$, represented as grey dots for readability). See text for details on the interaction potentials. (c) Typical snapshots: starting from the same initial configuration, an equilibrium simulation, where species interconversions only follow the passive pathway ($\Delta\mu=0$), leads to macroscopic phase separation, whereas a nonequilibrium simulation ($\Delta\mu >0$) leads to interrupted phase separation and selection of a finite droplet size.}
    \label{fig_model}
\end{figure}

From a thermodynamical point of view, the stability of finite-size condensates indicates that Oswald ripening is arrested: it was suggested that this could originate from the interplay between phase separation and nonequilibrium chemical reactions ~\cite{Zwicker2022}. More precisely, in models for active droplets, biomolecular condensates are typically made of proteins that coexist under two states: one in which they tend to form droplets, and one in which they do not. The conversions between these two state are assumed to break detailed balance. This idea, that dates back to the studies of nonequilibrium dissipative structures~\cite{Glotzer1994, Christensen1996}, has become prominent in recent theoretical studies and in the context of biomolecular condensates: Flory-Huggins-like continuous descriptions have been designed to account for the formation of active emulsions and chemically active droplets~\cite{Zwicker2015, Zwicker2017, Weber2019,Tjhung2018,Ziethen2023}, and can be engineered experimentally with coarcevates~\cite{Nakashima2021, Nakashima2018}.

However, so far, models of chemically active droplets are spatially continuous and deterministic, in such a way that they cannot account for the polydispersity in size and shape of the droplets. Moreover, these models cannot resolve the dynamics of the system in the stationary state, where fluctuations are responsible for the continuous nucleation of new droplets, and their coalescence. {Both these aspects are crucial at the mesoscale, i.e. when condensates are very small and where thermal fluctuations play a prominent role. Recently, in experiments, this appeared to be a particularly relevant level of description~\cite{Keber2024}. Additionally, such models rely on the numerical resolution of partial differential equations which are very costly and which then have been limited to 2D systems~\cite{Zwicker2015, Zwicker2017, Tjhung2018,Ziethen2023}.} In this context, it is mandatory to design particle-based simulations of chemically active droplets. This represents a challenge, both in the understanding of the relevant timescales and microscopic details of the model, and in the interpretation of the data, which requires advanced computational tools.

In this Letter, we propose a microscopic and stochastic model for chemically active droplets. Using Brownian dynamics simulations of   model proteins that switch between two states (Fig. \ref{fig_model}(a)) and that transiently attract each other (Fig. \ref{fig_model}(b)), we identify the conditions under which stable, finite-size droplets may form. The nonequilibrium ingredient of the model is a chemical drive, which breaks detailed balance, and whose amplitude controls the size of the droplets, their polydispersity, and their shape anisotropy (Fig. \ref{fig_model}(c)). In addition, we fully characterize the nonequilibrium steady state, by monitoring the coalescence and shrinkage processes at the level of individual droplets -- an aspect out of reach from previous deterministic, coarse-grained description. To our knowledge, the present work constitutes the first particle-based simulations of active droplets that form through the interplay between phase separation and chemical reactions which are driven away from equilibrium in a controlled way.

\emph{Model.---} We consider a three-dimensional suspension of Brownian particles made of three species, denoted by $A$, $B$ and $C$.  $A$ particles may convert into $B$, and vice versa -- the dynamics of these interconversions will be specified later on. To account for the high density of the intracellular medium \cite{Ellis2003,Zimmerman1993}, $C$ particles are crowders that do not undergo any reaction. Their density is chosen in such a way that the total volume fraction is~$0.1$.

We denote by $S_n(t)\in\{A,B,C\}$ the species of particle $n$ at time $t$. We assume that the positions of particles $\rr_1,\dots,\rr_N$ obey overdamped Langevin equations (see  Supplementary Material~\cite{SM} for details on the numerical methods).
The pair interaction between two particles $m$ and $n$, denoted by $U_{S_n S_m} (r_{mn})$, depends on their species and on their relative distance $r_{mn}=|\rr_m-\rr_n|$. 
The evolution equations of the particles positions read:
\begin{align}
    \frac{\dd \rr_n}{\dd t} = &\sqrt{2D} \boldsymbol{\eta}_n(t) -\frac{D}{\kB T} \sum_{m\neq n}\nabla U_{S_n, S_m}  (r_{mn}), 
    \label{overdampedLangevin}
\end{align}
where we assume that all the particles have the same bare diffusion coefficient $D$, and where $\boldsymbol{\eta}_n(t)$ is a Gaussian white noise of zero average and unit variance $\langle \eta_{n,i}(t)\eta_{m,j}(t') \rangle = \delta_{ij}\delta_{nm}\delta(t-t')$. Since the dynamics is overdamped, the velocities of the particles at a given time are irrelevant, and the state of the system is completely described by the configuration vector $\mathcal{C}=(\rr_1,\dots,\rr_N;S_1,\dots,S_N)$.
 The $B$ particles interact with each other through a Lennard-Jones (LJ) potential, which is truncated at a distance $r_c=2.5\sigma$ ($\sigma$ being the diameter of the particles), and shifted in order to ensure continuity of the potential at $r=r_c$. It reads $U_{B,B}(r)=[U_{\varepsilon}(r)-U_{\varepsilon}(r_c)]\theta(r_c-r)$, where $U_{\varepsilon}(r)=4\varepsilon\left[\left(\frac{\sigma}{r}\right)^{12} -   \left(\frac{\sigma}{r}\right)^6 \right]$ is the standard LJ potential and $\theta(r)$ denotes the Heaviside function. All the other pair interactions are purely repulsive and are modeled by the Weeks-Chandler-Andersen (WCA) potential~\cite{Weeks1971}, which is simply a  Lennard-Jones potential truncated and shifted at $r=2^{1/6}\sigma$: $U_{A,\{ A,B,C\}} = U_{C,\{ A,B,C\}} =[U_{\varepsilon'}(r)+\varepsilon'] \theta(2^{1/6}\sigma-r)$. The energy parameters of the interaction potentials are $\varepsilon'=\kB T$ and $\varepsilon=2 \kB T$:  the latter ensures that, in the absence of species interconversion, the $B$ particles phase separate at the considered density.  Throughout the paper, the distances will be measured in units of $\sigma$,  the energies in units of $\kB T$ and times in units of $\sigma^2/D$.

In order to specify the rules of species interconversions, let us consider two configurations $\mathcal{C}$ and $\mathcal{C}'$, which only differ by the species of one particle. We assume that the species of each particle obey a random telegraph process~\cite{Gardiner1985}, in such a way that the probability for the system to be in configuration $\mathcal{C}'$ at time $t + \delta t$ knowing that it was in configuration $\mathcal{C}$ at time  $t$ reads, for a sufficiently small $\delta t$: $ P(\mathcal{C}', t+\delta t |\mathcal{C}, t) \simeq k_{\mathcal{C},\mathcal{C}'} \delta t$, where $k_{\mathcal{C},\mathcal{C}'}$ is the rate at which the transition takes place. A crucial feature of our model is that the transition rates  have two contributions: a passive one $k^{\text{p}}_{\mathcal{C},\mathcal{C}'}$, and an active  one $k^{\text{a}}_{\mathcal{C},\mathcal{C}'}$, in such a way that $k_{\mathcal{C},\mathcal{C}'} = k^{\text{p}}_{\mathcal{C},\mathcal{C}'}+k^{\text{a}}_{\mathcal{C},\mathcal{C}'}$.

First, when the system is at equilibrium, interconversions take place through a `passive' pathway. The conversion rates must obey the detailed balance condition:   $ { k^{\text{p}}_{\cC',\cC}}/{k^{\text{p}}_{\cC,\cC'}} = \exp\{-\beta[E(\cC)-E(\cC')]\}$, where $\beta=(\kB T)^{-1}$, and where $E(\cC) = \frac{1}{2}\sum_{m\neq n }U_{S_nS_m}(r_{mn})+\sum_n w_{S_n}$, with $w_{S_n}$ being the internal energy of a particle of species $S_n$. Second, we consider the situation where detailed balance is broken, and where interconversions take place through an `active' pathway. In this situation, we write the ratio between the rates as ${ k^{\text{a}}_{\cC',\cC}}/{k^{\text{a}}_{\cC,\cC'}} = \exp\{-\beta[E(\cC)-E(\cC')+2\kappa_{\cC',\cC}\Delta \mu]\}$ where $\Delta \mu$ is a chemical drive, and $\kappa_{\cC',\cC}=1$ if the transition from $\cC'$ to $\cC$ implied the formation of a $B$ particles and $-1$ otherwise {(the reason for this choice will be made clear in the next paragraph)}. The active pathway can be interpreted as follows, if we consider for example the common involvement of ATP in biochemical reaction: if the formation of a $B$ particle is only possible by ATP consumption ($A+\text{ATP}\rightleftharpoons B+\text{ADP}$) and assuming that the chemical potentials of ATP and ADP are almost constant (i.e. they are chemostatted), the chemical drive is actually given by the difference between the chemical potentials of the chemostatted species: $\Delta \mu = \mu_{\rm ATP} - \mu_{\rm ADP}$~\cite{Julicher1997}.

To form stable droplets, {and to avoid the formation of a single one,} the formation of $A$ particles must be favored in dense regions (i.e. where droplets tend to form), and, on the contrary, the formation of $B$ particles must be favored in dilute regions. Following the idea from Refs.~\cite{Zwicker2022,Zwicker2017,Weber2019}, we assume that $ k^{\text{p}}_{\cC',\cC} = k_0 (1-\phi_\text{loc}/\phi_\text{max}) \ex{-\frac{\beta}{2}[E(\mathcal{C})-E(\mathcal{C}')]}$  and $ k^{\text{a}}_{\cC',\cC} = k_0 (\phi_\text{loc}/\phi_\text{max}) \ex{-\frac{\beta}{2}[E(\mathcal{C})-E(\mathcal{C}')+2\kappa_{\cC',\cC}\Delta \mu]}$, where $\phi_\text{loc}$ is the local density of $A$ and $B$ particles around the particle whose species change between configuration $\mathcal{C}$ and $\mathcal{C'}$, and where $\phi_\text{max}$ is the maximum  volume fraction of the mixture and is approximated to the maximum packing fraction in 3 dimension: $\phi_\text{max}\simeq 0.74$.  We fix  $k_0=10^{-2}$ in all the simulations. With this choice of the reaction rates $k^\text{a,p}_{\cC',\cC}$, and at equilibrium ($\Delta \mu = 0$), global detailed balance is fulfilled: $k_{\cC',\cC}/k_{\cC,\cC'} = \exp\{-\beta[E(\cC)-E(\cC')]\}$. Therefore, the parameter $\Delta \mu$ controls finely the deviation from equilibrium, as opposed to previous models of mixtures
 of particles with `active switching', which have been studied in other contexts~\cite{Decayeux2021a, Decayeux2022, Decayeux2023,Alston2022, Bley2021, Goth2022, Bley2022,Moncho-Jorda2020,Longo2022}. In those descriptions, the rates of conversion of the particles are independent of their local environment, in such a way that the distance of the system to equilibrium is difficult to evaluate.
 
{Another key parameter of our model is the difference between the internal energies $\Delta w = w_B-w_A$. Here, we take $w_A=w_C=0$ and $w_B=0.5$, meaning that $\Delta w>0$, and that $A$ is more stable than $B$ in the absence of interactions. However, stable droplets can also be formed for other choices of $\Delta w$ (including $\Delta w<0$). This simply requires to explore other values of the chemical drive $\Delta \mu$. The choice of parameters is discussed in details in SM, and we show that the formation of stable droplets is not restricted to the parameters shown in the main text \cite{SM}.}

\begin{figure}
    \centering
    \includegraphics[width=\columnwidth]{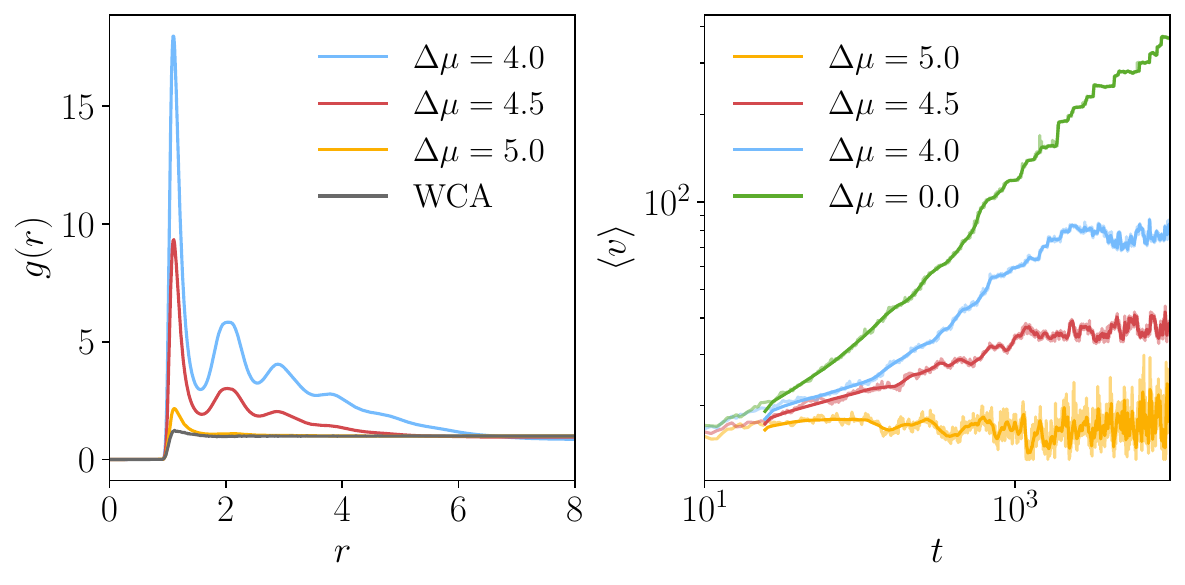}
    \caption{{ Left: Radial distribution function (RDF) between $\{A, B\}$ particles in the stationary state (for $t$ between $10^4$ and $6\cdot 10^4$). The RDF for particles in a system at the same density and interacting via a WCA potential is shown in grey for reference.} Left: Average volume of the droplets  $\langle v\rangle$  as a function of time for  various values of the chemical drive $\Delta \mu$. Solid lines are running averages (over 50 points) of the raw data, shown in lighter color. }
    \label{fig_v_t}
\end{figure}

\emph{Formation of active droplets.---}  We perform numerical simulations and tune the parameter $\Delta\mu$, which quantifies the deviation from thermal equilibrium, and which will be varied from 0 to 5. In the range of parameters we consider, we observe that the $B$ particles tend to form droplets (Fig. \ref{fig_model}(b,c) and Supplementary Movies). 
{We compute the radial distribution functions of $A$ and $B$ particles, in the stationary state and for $\Delta \mu >0$ (Fig. \ref{fig_v_t}). These functions bear the signature of the formation of dense clusters, that become less structured as $\Delta \mu$ increases. In order to characterize quantitatively the spatial structure, we perform a cluster analysis on each trajectory~\cite{SM}, and compute the average volume of the droplets at time $t$, denoted by $\langle v\rangle $.}

At equilibrium, i.e. when $\Delta \mu = 0$, we observe that $ \langle v\rangle$ increases as a power law ($ \langle v\rangle \sim t^\alpha$, with $\alpha\simeq0.48$, see Fig. \ref{fig_v_t}). This slow increase is characteristic of the progressive coarsening of a single droplet, as expected for purely passive systems, through Ostwald ripening. On the contrary, when $\Delta \mu >0$, one observes that $ \langle v\rangle $  quickly saturates at a finite value, that depends on $\Delta \mu$, and that is much smaller than the typical volumes reached at equilibrium { (see Supplementary Movie for examples of simulation trajectories). This defines a stationary state in which, although $\langle v \rangle$ does not vary anymore, droplets coalesce and are continuously nucleated in the dilute phase}.  We then conclude that the active reaction pathway results in droplet size selection, as predicted by continuous-space reaction-diffusion theories \cite{Zwicker2015,Zwicker2017,Zwicker2022}. Note that for values of $\Delta\mu$ smaller than 4, the typical droplet volume that is selected by the active reaction pathway is large, and is likely to be comparable to system size, in such a way that strong finite size effects prevent us from reaching stationary state in a reasonable computational time. 

\begin{figure}
    \centering
    \includegraphics[width = \columnwidth]{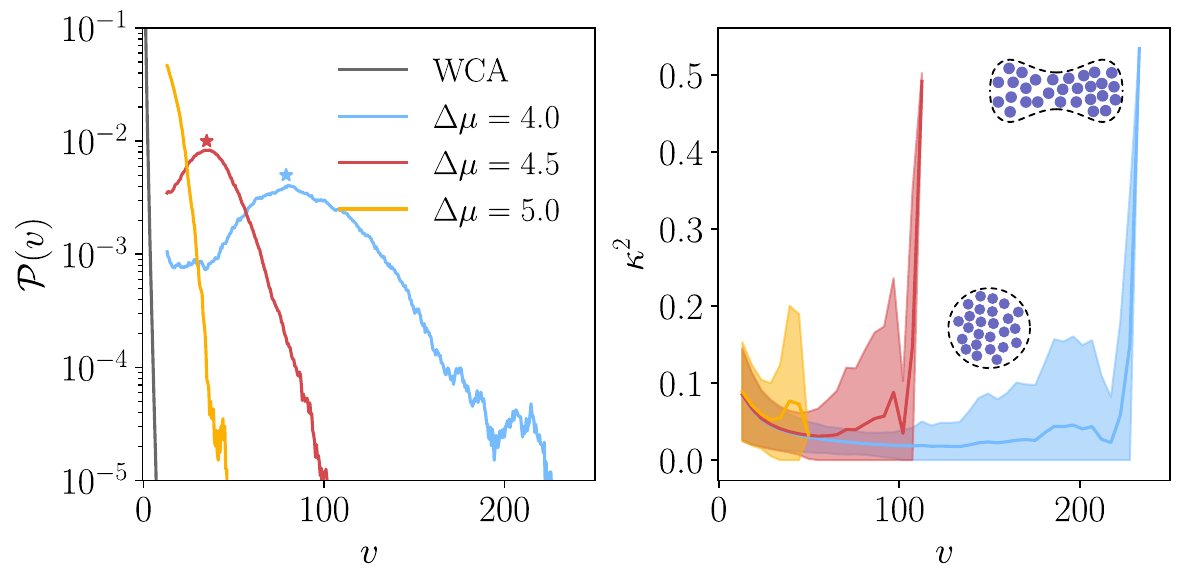}
    \caption{Left: Probability distribution of the droplet volume $v$, for various values of the chemical drive $\Delta \mu$, in the stationary state. { The two stars represent, for $\Delta \mu = 4 $ and $4.5$, the stable fixed points in the phase portrait shown on Figure~\ref{fig_DeltaR}}. Right: Shape anisotropy of the droplets as a function of their volume $v$. Solid lines are running averages (over 10 points), and the widths of the colored areas are the standard deviations (insets: schematics of the spherical and non-spherical droplets, which correspond respectively to low and high values of $\kappa^2$).}
    \label{fig_kappacount}
\end{figure}

\emph{Polydispersity and shape anisotropy of the active droplets.---}  As seen on Fig.~\ref{fig_v_t}, the volume of the droplets strongly fluctuates around its average value. To quantify the polydispersity of the droplets, we measure the volume of each droplet at each time step and in the stationary state, and we plot the resulting histograms on Fig.~\ref{fig_kappacount}. For small enough values of $\Delta \mu$, the distributions typically show a wide peak at a large value of $v$, which represents the volume of the droplet selected by the active pathway. 
Furthermore, the distribution of $v$ is large, which shows that the droplets are very polydisperse.  This is in contrast with purely deterministic approaches, which result in the selection of a single droplet size, with vanishing fluctuations. Finally, we observe that the variances of these distributions decrease as the chemical drive increases.

{ The probability distribution $\mathcal{P}(v)$ does not contain information about the shapes of the droplets. We characterize them through the computation of the gyration tensors of the droplets and their eigenvalues \cite{SM}.} We introduce a coefficient $\kappa^2$, called shape anisotropy, and that varies between $0$ and $1$, which correspond respectively to a perfectly spherical droplet and a perfectly elongated one. As seen on Fig.~\ref{fig_kappacount}, for small enough values of $v$, the shape anisotropy $\kappa^2$ decreases with $v$: this is expected, as the $B$ particles constituting the droplets arrange in such a way to minimize surface tension. However, for larger values of $v$, we observe that the shape anisotropy may become an increasing function of the droplets volume. This means that, in this range of parameters, droplets tend to have a more elongated shape than the one that minimizes their surface tension. We interpret this as follows: under the effect of thermal fluctuations, droplets diffuse in the simulation box and coalesce, thus forming large droplets whose shape is elongated (see inset of Fig~\ref{fig_kappacount} { and Supplementary Movie}). 
Under the combined effect of surface tension and active chemical reactions, droplets that are formed by coalescence events relax to smaller and more spherical droplets. { Note that large, aspherical droplets, that result from coalescence events, are quite rare in our simulations and correspond to the tails of the distributions shown on Fig~\ref{fig_kappacount}, left.}

\begin{figure}
    \centering
    \subfigure{\raisebox{5mm}{\includegraphics[width = 0.3\columnwidth]{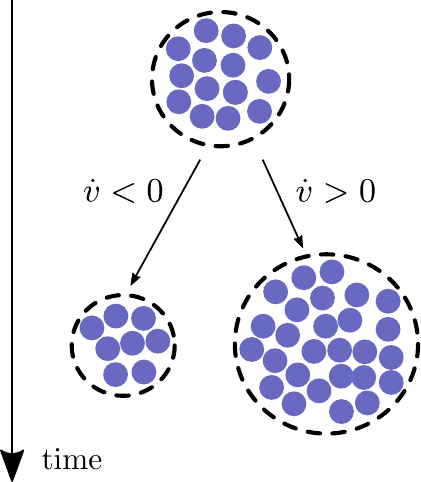}}}
    \subfigure{\includegraphics[width = 0.59\columnwidth]{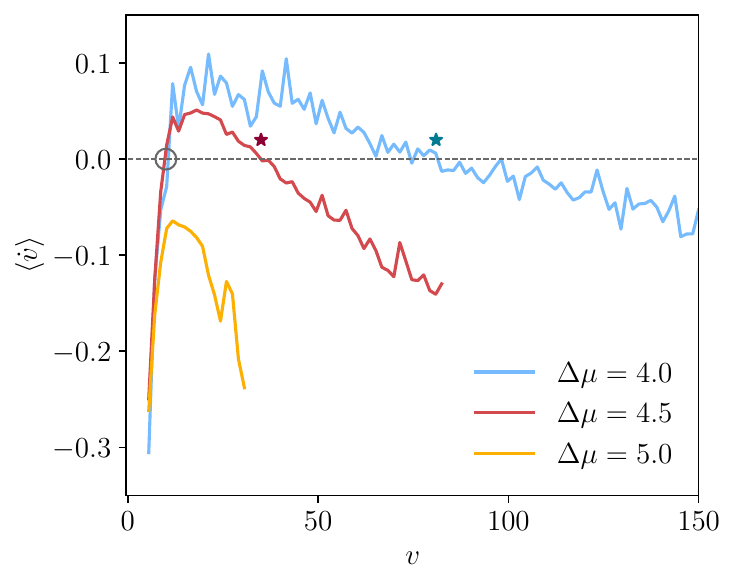}}
    \caption{ { Left: Schematic representation of $\dot{v}$. Right: Phase portrait of the volume of droplets (derivative of the volume with respect to time as a function of the volume) for different values of the chemical drive $\Delta\mu$. For $\Delta\mu = 4$ and $4.5$, the two stars correspond to the typical volume reached by a droplet (shown on Fig. \ref{fig_kappacount}). Open symbol: unstable fixed point, stars: stable fixed points.}}
    \label{fig_DeltaR}
\end{figure}

\emph{Size selection mechanism and lifetime of the active droplets.---} In order to support this claim, we now study the time evolution of the droplets in the stationary state. Indeed, so far, we have only identified droplets at each timestep without linking them from one step to another. To go further, we design an algorithm to track droplets in time \cite{SM}. The volume of each droplet is monitored from its birth (that may occur from spontaneous nucleation) to its death (that may occur through shrinking below a threshold value, or coalescence with another droplet). For each droplet that is then identified and tracked, we compute the derivative of $v$ with respect to time and plot the resulting phase portrait (the average $\langle \dot{v} \rangle$ as a function of $v$) on Fig~\ref{fig_DeltaR}.

This plot shows that, for values of the chemical drive that lead to the selection of large droplets ($\Delta \mu = 4$ or $4.5$), the curves cross the horizontal axis twice: there are then two fixed points. First, at small volumes, the unstable fixed point represented by the empty circle on Fig~\ref{fig_DeltaR}, corresponds to the critical nucleation volume $v_{\rm crit}$: below this volume, droplets are unstable and dissolve spontaneously.
{ Second, there is a stable fixed point at a volume that we will call $v^*$ and that is represented by the two stars on Fig.~\ref{fig_DeltaR}. We report the values of $v^*$ on the probability distribution shown on Figure~\ref{fig_kappacount} and observe that they correspond to the maxima of $\mathcal{P}(v)$.}
More precisely, when the droplets have initially a volume between $v_{\rm crit}$ and $v^*$, the derivative of their volume with respect to time is positive, meaning that they tend to grow by aggregation of $B$ particles from the dilute phase, or coalescence with other droplets. On the contrary, when their initial volume is greater than $v^*$, $\langle \dot{v}\rangle$ is negative, meaning that they shrink until they reach the stable value $v^*$.  

This mechanism for droplet size selection is therefore a microscopic and stochastic counterpart to the predictions from continuous and deterministic models  \cite{Zwicker2015}. In such theories, the typical volume reached by the droplets at equilibrium is understood as a compromise between a growth (resp. loss) term, which dominates at small (resp. large) volumes. However, as opposed to the ideal case considered in previous theories (droplets in infinite volume without fluctuations), we cannot identify these terms unambiguously given the limited statistics at small volumes.

\begin{figure}
    \centering
    \includegraphics[width=\columnwidth]{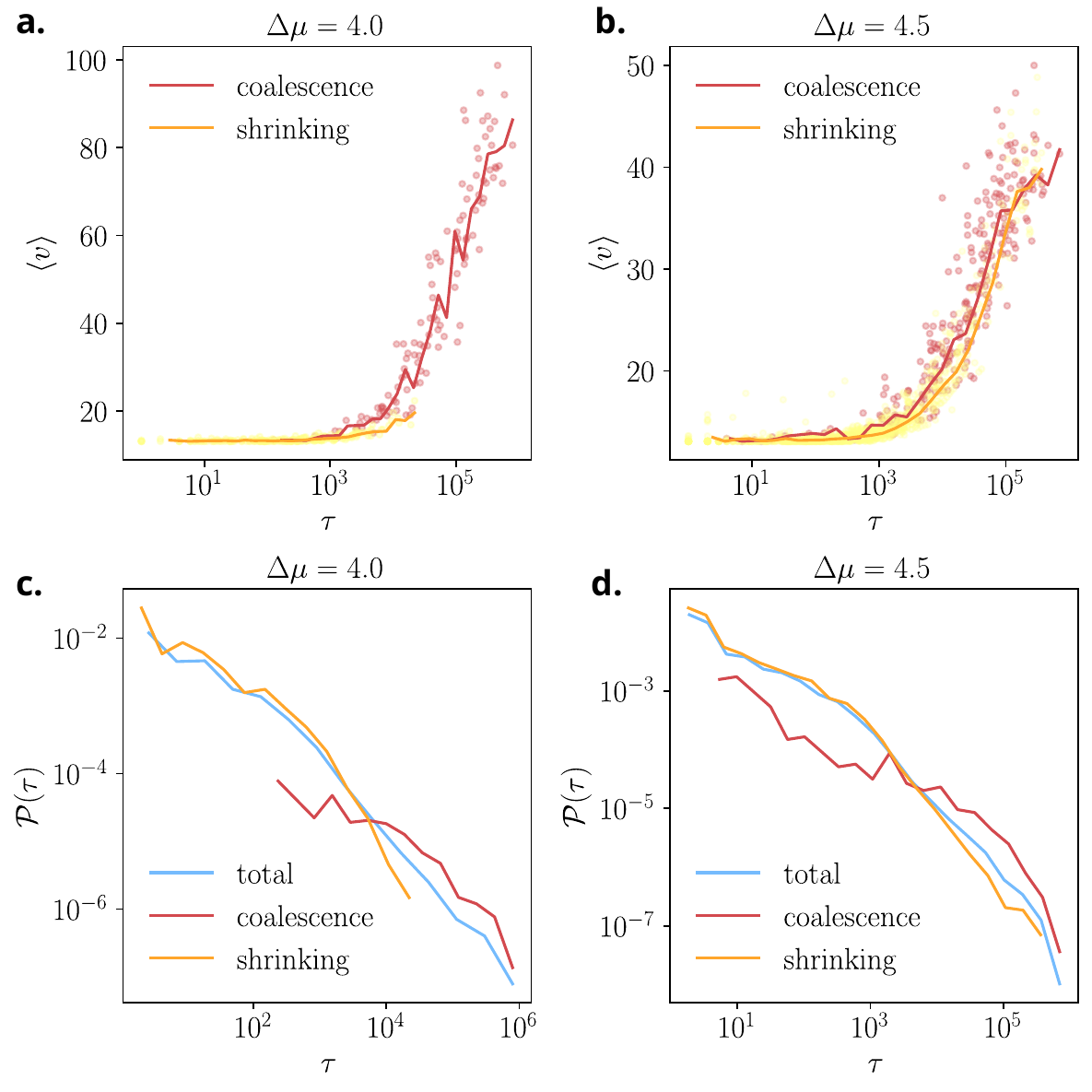}
    \caption{{ Average volume of the droplet during its life as a function of the lifetime $\tau$ for different death events and for $\Delta\mu =4.0$ (a) and $\Delta\mu =4.5$ (b). Probability distribution of droplet lifetime $\tau$ for different death events and for $\Delta\mu =4.0$ (c) and $\Delta\mu =4.5$ (d).}}
    \label{fig_histo_tdv2}
\end{figure}

Finally, we show that our stochastic simulations capture the broad range of behaviors that emerge in the nonequilibrium steady state. To this end, for each droplet identified and tracked in the trajectories, we compute the number of steps elapsed between its birth and its death~\cite{SM} -- this `lifetime' will be denoted by $\tau$. { We observe that there are two reasons for droplets to disappear: they either coalesce with another one or shrink below the critical volume $v_\text{crit}$. We show on Figure~\ref{fig_histo_tdv2}(a,b) the average volume of the droplet during their life as function of their lifetime : the data is split depending on the type of death event. For $\Delta\mu = 4$, large droplets (of volume close to $v^{*}$) are very unlikely to shrink below the critical volume $v_\text{crit}$: this is consistent with the phase portrait where $v^{*} \gg v_\text{crit}$. As a consequence, they mostly experience coalescence events and their lifetime is much larger then the one of smaller droplets as can be seen on the probability distribution of $\tau$, shown on Figure~\ref{fig_histo_tdv2}(c). On the contrary, for $\Delta\mu = 4.5$, large droplets can either coalesce or shrink: since $v^{*}$ is closer to $v_\text{crit}$ fluctuations can easily bring any droplet below the critical nucleation volume. In this case, the lifetime depends much less on the size of the droplet (Fig.~\ref{fig_histo_tdv2}(b,d)).} Interestingly, this shows that the active pathway controls not only the size and shape of the droplets, but also the time during which they persist and keep $B$ particles in close vicinity. This property may play a key role in some biological processes, where condensates are known to act as microscopic `reactors', that bypass the diffusion-limited step of chemical reactions \cite{Strulson2012,Castellana2014}.

\emph{Conclusion.---} In this Letter, we introduced an unprecedented particle-based computational framework  designed to study active droplets. Our microscopic and stochastic approach enabled us to simulate Brownian suspensions where Ostwald ripening is interrupted in a controlled way, allowing for a detailed characterization of the polydispersity, shape anisotropy, and fluctuations of active droplets in the nonequilibrium steady state of the system. Our work not only advances the understanding of active emulsions but also provides a versatile methodology applicable to a wide range of systems. As a perspective, the present work opens the way to the understanding of the effect of macromolecular crowding on the formation of biomolecular condensate \cite{Andre2020,Spruijt2023}.

\emph{Acknowledgments.---}  The authors thank David Zwicker and Chengjie Luo for discussions.


%

\end{document}


\beginsupplement

\title{Microscopic and stochastic simulations of chemically active droplets}

\author{Roxanne Berthin}
\thanks{These two authors contributed equally.}
\affiliation{Sorbonne Universit\'e, CNRS, Physico-Chimie des \'Electrolytes et Nanosyst\`emes Interfaciaux (PHENIX), 4 Place Jussieu, 75005 Paris, France}

\author{Jacques Fries}
\thanks{These two authors contributed equally.}
\affiliation{Sorbonne Universit\'e, CNRS, Physico-Chimie des \'Electrolytes et Nanosyst\`emes Interfaciaux (PHENIX), 4 Place Jussieu, 75005 Paris, France}

\author{Marie Jardat}
\affiliation{Sorbonne Universit\'e, CNRS, Physico-Chimie des \'Electrolytes et Nanosyst\`emes Interfaciaux (PHENIX), 4 Place Jussieu, 75005 Paris, France}

\author{Vincent Dahirel}
\affiliation{Sorbonne Universit\'e, CNRS, Physico-Chimie des \'Electrolytes et Nanosyst\`emes Interfaciaux (PHENIX), 4 Place Jussieu, 75005 Paris, France}

\author{Pierre Illien}
\affiliation{Sorbonne Universit\'e, CNRS, Physico-Chimie des \'Electrolytes et Nanosyst\`emes Interfaciaux (PHENIX), 4 Place Jussieu, 75005 Paris, France}

\maketitle

\onecolumngrid

\begin{center}
\textbf{\textit{\large Supplemental Material}}
\end{center}

\tableofcontents

\section{Numerical simulations}
\label{app_numsim}

To perform Brownian dynamics simulations, we use the LAMMPS computational package~\cite{Thompson2022}, with the command ‘fix brownian’. This command integrates the evolution equations of the positions of particles (Eq. (1) in the main text) thanks to a forward Euler-Maruyama scheme~\cite{Allen1987}:
\begin{equation}
   \rr_n(t+\delta t) =\rr_n(t) + \sqrt{2D \delta t}  \boldsymbol{\xi} -\delta t\frac{D}{\kB T} \sum_{m\neq n}\nabla U_{S_n, S_m}  (r_{mn}).
    \label{overdampedLangevin_disc}
\end{equation}
The timestep $\delta t$ is  equal to $2\cdot 10^{-4}$, and  $\boldsymbol{\xi}$ is a random vector drawn from a normal distribution of mean $0$ and unit variance.

To implement species interconversions,  we  use PyLammps, the wrapper Python class for LAMMPS. With the command `find\_pair\_neighlist', for each reacting particle ($A$ or $B$), and at each time step, we get the number $N_\text{loc}$ of particles that are located at a distance smaller than $2.5\sigma$ (i.e. the cutoff of the interaction potential).  The local volume fraction $\phi_\text{loc}$, used to compute the reaction rates $k_{\mathcal{C}',\mathcal{C}}^\text{a,p}$ (see main text) is computed as $N_\text{loc}/V_\text{loc}$, where $V_\text{loc}$ is the average of the volumes of two spheres of radius $2.5\sigma$ and $3\sigma$ (as particles at a distance greater than $2\sigma$ have a part of their volume in the corona between those two spheres). We then compute the difference of energy of the configuration before and after the interconversion, which is accepted with the probabilities specified in the main text. For computational efficiency, the species interconversion are evaluated each $100 \delta t$.

The simulations are performed in a cubic box of length $27.14$ with periodic boundary conditions. The particles are initially located on a face-centered cubic lattice. This initial configuration is equilibrated for $10^5$ timesteps. The initial number of each species is  $N_A^0=200$, $N_B^0=800$ and $N_C^0=2820$.  For each set of parameters, we carried out 10 different runs with different initial conditions, and different seeds in the Brownian dynamics command ‘fix brownian’.

\section{Data analysis}
\label{app_data_analysis}

\subsection{Cluster analysis}

To have information about the spatial distribution of the droplets, we first need to identify the droplets at each timestep. To do so, we perform a cluster analysis of the trajectories. 
We work only with the $B$ particles (that will form the droplets) and store their positions every {200 $\delta t$ (for droplet tracking) and 5000 $\delta t$ (for average droplet size)}. We then {search for the neighboring particles in each stored step with a cell list algorithm and compute all the distances between neighboring particles}. Then, if two $B$ particles are at a distance $d \le 1.5\sigma$, we assume that they belong to the same droplet. This value corresponds to the first peak of the $B - B$ particles radial distribution function (shown on the right panel of Figure~2) and thus corresponds to the typical distance between $B$ particles in a droplet. {Moreover, except for building Fig. 4, we only considered droplets which are made of at least 25 particles: this corresponds to the critical nucleation volume $v_\text{crit}$}.

We then store, at each step, the positions and labels of $B$ particles into an array of sub-arrays. We then get an array in which each sub-array corresponds to a droplet and contains the labels of the particles that constitute it.
From this, we are able to compute { the average volume of the droplets and }the volume of each droplet, {which is defined as the sum of the volume of each one of its components. This yields for $\mathcal{N}$ particles : $\mathcal{N}\cdot\frac{4}{3}\pi(\frac{\sigma}{2})^3$. We then average those volumes in Fig. 2, 3, 4 and 5 of the main text .

To compute the relative shape anisotropy in Fig.~3, we start by deriving the gyration tensor $S_{mn}$ which is a 3$\times$3 matrix describing the second moments of positions of all the particles that belong to a given droplet of $\mathcal{N}$ particles:
\begin{equation}
    S_{mn} = \frac{1}{2 \mathcal{N}^2} \sum_{i,j \in\text{droplet}} (r_{m,i} - r_{m,j})  (r_{n,i} - r_{n,j}),
\end{equation}
with $r_{m,i}$ the $m^{\rm th}$ Cartesian coordinate of the position  $\rr_i$ of the $i^{\rm th}$ particle.  The eigenvalues of matrix ${\bf S}$ are denoted by $\lambda_{n}^2$. From them, we get the relative shape anisotropy defined as : 
\begin{equation}
    \kappa^2=\frac{3}{2}\frac{\sum_n\lambda_{n}^{4}}{(\sum_n\lambda_{n}^{2})^2} - \frac{1}{2}.
\end{equation}
With this definition, $\kappa^2$ is comprised between 0 and 1. When $\kappa^2$ = 1, all the particles are arranged along a line, whereas $\kappa^2$ = 0 means that all particles are arranged in a perfect sphere.}
 
\subsection{Droplet tracking}

Once we have identified each droplet at each step, we want to track them during time, by linking them between consecutive steps to another. 

To do so, for each pair of consecutive timesteps, we construct the tracking matrix $\mathbf{T}$, of dimension $M\times N$, with $M$ the number of droplets at step $i$ and $N$ the number of droplets at step $i+1$. Each element $T_{mn}$ corresponds to a percentage of similarity between the two droplets $m$ and $n$. To compute $T_{mn}$, we compare the labels of the $B$ particles that belong to the droplet $m$ of step $i$, to the labels of the $B$ particles that belong to the droplet $n$ of step $i+1$. Each time a label is the same in droplet $m$ and droplet $n$, $T_{mn}$ is incremented by 1. We finally divide $T_{mn}$ by $\text{min}(M,N)$ and we assume that two droplets are linked from step $i$ to step $i+1$ if $T_{mn} \ge 0.5$. This criterion results in the following rules:

{
\begin{itemize}
    \item If no element in row $m$ is $\ge 0.5$, droplet indexed $m$ at time $t$ has no descendant at time $t+1$ and is considered dead by shrinking.
    \item If no element in column $n$ is $\ge 0.5$, droplet indexed $n$ at time $t+1$ has no ancestor at time $t$ and is considered as a newly nucleated droplet.
    \item If row $m$ and column $n$ have only one element $\ge 0.5$ and this element is shared, the droplet indexed $m$ at $t$ is linked to the droplet indexed $n$ at $t+1$.
    \item If two elements $T_{mn}$ and $T_{mn'}$ of row $m$ are $\ge 0.5$, this corresponds to the separation of the mother droplet indexed $m$ at time $t$ in two daughter droplets indexed $n$ and $n'$ at time $t+1$. The mother droplet is linked to the biggest daughter droplet and the smallest daughter droplet is considered as a new droplet.
    \item If two elements $T_{mn}$ and $T_{m'n}$ of column $n$ are $\ge 0.5$, this corresponds to the fusion of two mother droplets indexed $m$ and $m'$ at time $t$ in one daughter droplet indexed $n$ at time $t+1$. The biggest mother droplet is linked to the daughter droplet and the smallest mother droplet is considered as dead by coalescence.
\end{itemize}}

{We track all the droplets at each step and count the number of steps during which they exist. The droplets for which we did not see both birth and death are ignored. }

{

\section{Supplementary Results}

\begin{enumerate}

    \item \textbf{Finite size effects:} We show on Figure~\ref{fig:box_size_effect} the probability distribution of droplet volume $\mathcal{P}(v)$ for various value of $\Delta\mu$ and for two system sizes. The plot shows that the distribution $\mathcal{P}(v)$ is not significantly affected by the size of the simulation box. 
    
\begin{figure}
    \centering
    \includegraphics[scale=0.6]{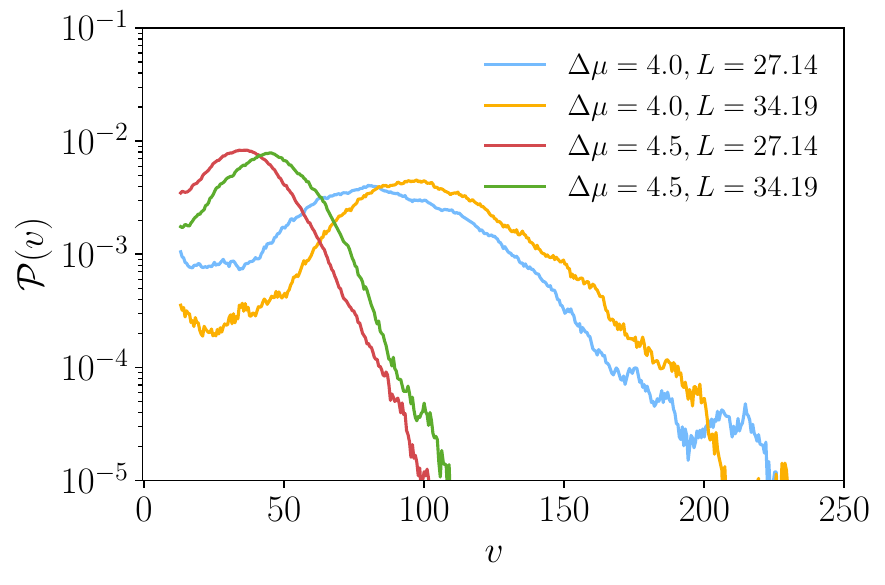}
    \caption{Probability distribution of droplet volume $\mathcal{P}(v)$ for various value of $\Delta\mu$ and for two system sizes.}
    \label{fig:box_size_effect}
\end{figure} 

    \item \textbf{Polydispersity:} On Fig. \ref{fig:distribution_width}, we show the probability distribution of the droplet volume for two reference cases: the case with equilibrium reactions ($\Delta\mu=0$), and the case where all the particles are of type WCA. We also compute the standard deviation of the distribution for different values of $\Delta \mu$ (Table~\ref{tab:std}).

\begin{figure}
    \centering
    \includegraphics[scale=0.8]{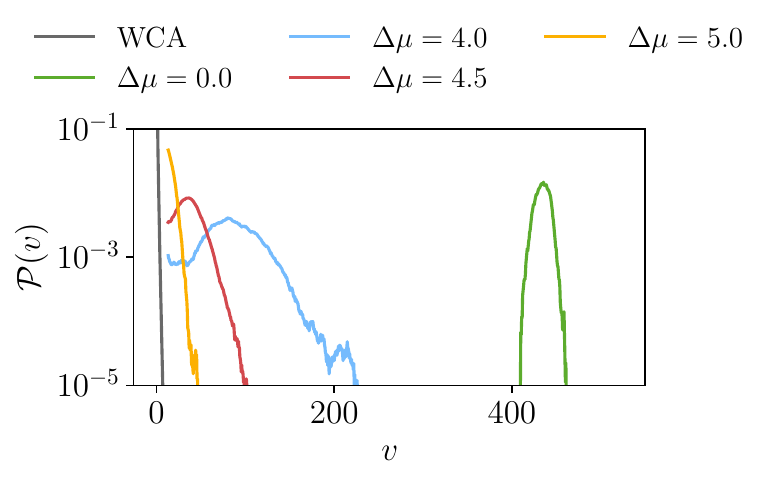}
    \caption{Probability distribution of the droplet volume $v$, for various values of the chemical drive $\Delta \mu$, ranging from 0 to 5, in the stationary state.}
    \label{fig:distribution_width}
\end{figure}

\begin{table}[t]
    \centering
    \begin{tabular}{c | c}
    $\Delta\mu $ &  std($v$) / $\langle v \rangle$ \\
    \hline
    0.0 & 0.015 \\
    4.0 & 0.375 \\
    4.5 & 0.363 \\
    5.0 & 0.219 \\
    \hline
    \end{tabular}
    \caption{Standard deviation of $v$ divided by its average value for different values of $\Delta\mu$.}
    \label{tab:std}
\end{table}

\item \textbf{Evolution of phase composition with chemical drive:} We show in Table~\ref{tab:composition} the influence of $\Delta \mu$ on the composition of the dilute phase. 
We find that the dilute phase gets more concentrated in both $A$ and $B$ particles as $\Delta \mu$ increases. 

\begin{table}[t]
    \centering
    \begin{tabular}{c | c | c}
    $\Delta\mu $ &  $\phi_A$ & $\phi_B$\\
    \hline
    0.0 & $3.59$ $10^{-3}$ & $1.05$ $10^{-3}$ \\
    4.0 & $1.21$ $10^{-2}$ & $4.03$ $10^{-3}$\\
    4.5 & $1.61$ $10^{-2}$ & $4.61$ $10^{-3}$ \\
    5.0 & $2.03$ $10^{-2}$ & $5.03$ $10^{-3}$ \\
    \hline
    \end{tabular}
    \caption{Volume fraction of $A$ and $B$ particles in the dilute phase, as a function of the chemical drive $\Delta \mu$.}
    \label{tab:composition}
\end{table}

\end{enumerate}

\section{Supplementary Movie}

Brownian dynamics simulations of systems composed of particles labeled as $A$ (red), $B$ (blue) and $C$ (black dots), at two different values of $\Delta\mu$. The two initial configurations are identical. Depending on the value of $\Delta\mu$, one can observe two distinct behaviors. When $\Delta\mu$ is equal to 0 (left movie) we typically observe what happens at equilibrium, where a single droplet is formed at the end of the simulation, indicative of the Ostwald ripening mechanism. Conversely, when the chemical drive $\Delta\mu$ is greater than 0 (right movie, where $\Delta\mu =4$), the ripening mechanism seems to be partially interrupted, leading to the formation of multiple droplets. This active emulsion reaches a nonequilibrium steady state, characterized by frequent nucleation and coalescence events. The total duration of the trajectories is $2.5\cdot 10^7$ timesteps.

{

\section{Choice of parameters}

In this Section, we discuss the choice of parameters that may lead to the formation of stable active droplets. We emphasize that the rates of interconversions actually depend on three energy parameters:
\begin{itemize}
    \item the energy parameter of the Lennard-Jones potential $\varepsilon$: it is chosen in such a way that, in the absence of reactions and at the considered density, the $B$ particles phase separate;
    \item the internal energies $w_A$, $w_B$ and $w_C$. Since $C$ particles do not react, the choice of $w_C$ does not matter. Then, we only discuss the difference of internal energies $\Delta w=w_B-w_A$;
    \item the chemical drive $\Delta \mu$.
\end{itemize}

These energy parameters are involved in two distinct reactions:
\begin{itemize}
    \item the passive reaction, which takes place essentially in the dilute phase, and which aims at feeding droplets with the transformation of $A$ particles into $B$. The rates of the passive reaction are essentially controlled by the difference of internal energies $\Delta w$. Note that, if $\Delta w >0$ (resp $\Delta w <0$), the $A$ (resp $B$) particles are more stable than the $B$ (resp $A$) particles in the absence of interactions.
    
    \item the active reaction, which takes place essentially in the dense phase, and which aims at shrinking droplets with the transformation of $B$ particles into $A$. The rates of the active reactions depend on the three energy parameters $\varepsilon$ (since the particles are very close one to another in the dense phase, their pair interactions matter), $\Delta w$, and $\Delta \mu$.

\end{itemize}

\begin{figure}
    \centering
    \includegraphics[width=\columnwidth]{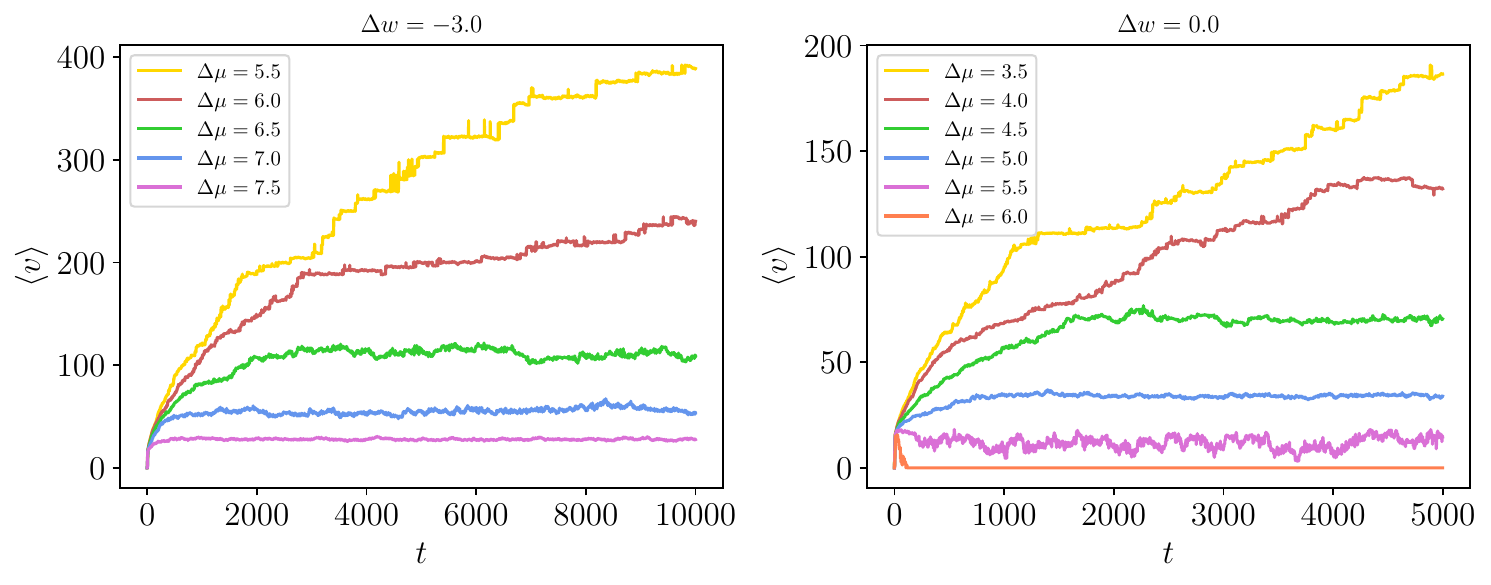}
    \caption{Left (resp. right): Average volume (resp. size) of the droplets  $\langle v\rangle$ (resp. $\langle \mathcal{N} \rangle$) as a function of time for various values of the chemical drive $\Delta \mu$, for two different variation of internal energies $\Delta w$.}
    \label{fig_supp_other_deltaw}
\end{figure}
    
The role of the chemical drive is to favor the formation of $A$ particles, and to go against the  mechanisms that stabilize the $B$ state, namely the Lennard-Jones interactions, and, when $\Delta w<0$, the difference in internal energies. We then expect that, when $\Delta w$ becomes negative and increases in absolute value, the chemical drive $\Delta \mu$ needed to stabilize the droplets should be larger. This is what we observe on Fig. \ref{fig_supp_other_deltaw}. For instance,  for $\Delta w<0$ (left panel), the range of $\Delta\mu$ that leads to stable droplets is $\Delta\mu \sim 6-7 \kB T$. On the contrary, as shown on the right panel ($\Delta w=0$) and in the main text ($\Delta w=0.5$), the values of $\Delta \mu$ that allow the formation of stable droplets are smaller when $\Delta w$ increases. In summary, the formation of stable droplets is very sensitive to the value of the difference in internal energies $\Delta w$.

}

%